\documentclass[twocolumn]{aastex631}
\usepackage{color}

\begin{document}

\title{CWISE J014611.20-050850.0AB: The Widest Known  Brown Dwarf Binary in the Field}

\author[0000-0002-1420-1837]{Emma Softich}
\affil{School of Earth and Space Exploration, Arizona State University, Tempe, AZ 85282, USA}

\author[0000-0002-6294-5937]{Adam C. Schneider}
\affil{United States Naval Observatory, Flagstaff Station, 10391 West Naval Observatory Rd., Flagstaff, AZ 86005, USA}
\affil{Department of Physics and Astronomy, George Mason University, MS3F3, 4400 University Drive, Fairfax, VA 22030, USA}

\author[0000-0001-9004-803X]{Jennifer Patience}
\affil{School of Earth and Space Exploration, Arizona State University, Tempe, AZ 85282, USA}

\author[0000-0002-6523-9536]{Adam J. Burgasser}
\affil{Center for Astrophysics and Space Science, University of California San Diego, La Jolla, CA 92093, USA}

\author[0000-0002-7260-5821]{Evgenya Shkolnik}
\affil{School of Earth and Space Exploration, Arizona State University, Tempe, AZ 85282, USA}

\author[0000-0001-6251-0573]{Jacqueline K. Faherty}
\affil{Department of Astrophysics, American Museum of Natural History, Central Park West at 79th St., New York, NY 10024, USA}

\author[0000-0001-7896-5791]{Dan Caselden}
\affil{Department of Astrophysics, American Museum of Natural History, Central Park West at 79th St., New York, NY 10024, USA}

\author[0000-0002-1125-7384]{Aaron M. Meisner}
\affil{NSF's National Optical-Infrared Astronomy Research Laboratory, 950 N. Cherry Ave., Tucson, AZ 85719, USA}

\author[0000-0003-4269-260X]{J. Davy Kirkpatrick}
\affil{IPAC, Mail Code 100-22, Caltech, 1200 E. California Blvd., Pasadena, CA 91125, USA}

\author[0000-0002-2387-5489]{Marc J. Kuchner}
\affil{Exoplanets and Stellar Astrophysics Laboratory, NASA Goddard Space Flight Center, 8800 Greenbelt Road, Greenbelt, MD 20771, USA}

\author[0000-0002-2592-9612]{Jonathan Gagn\'e}
\affil{Plan\'etarium Rio Tinto Alcan, Espace pour la Vie, 4801 ave. Pierre-de Coubertin, Montr\'eal, QC H1V~3V4, Canada}
\affil{Institute for Research on Exoplanets, Universit\'e de Montr\'eal, 2900 Boulevard \'Edouard-Montpetit Montr\'eal, QC H3T~1J4, Canada}

\author[0000-0001-8170-7072]{Daniella Bardalez Gagliuffi}
\affil{Department of Astrophysics, American Museum of Natural History, Central Park West at 79th Street, NY 10024, USA}

\author[0000-0001-7780-3352]{Michael C. Cushing}
\affil{Ritter Astrophysical Research Center, Department of Physics and Astronomy, University of Toledo, 2801 W. Bancroft St., Toledo, OH 43606, USA}

\author[0000-0003-2478-0120]{Sarah L. Casewell}
\affiliation{Department of Physics and Astronomy, University of Leicester, University Road, Leicester LE1 7RH, UK}

\author[0000-0003-2094-9128]{Christian Aganze}
\affil{Center for Astrophysics and Space Science, University of California San Diego, La Jolla, CA 92093, USA}

\author[0000-0002-5370-7494]{Chih-Chun Hsu}
\affil{Center for Astrophysics and Space Science, University of California San Diego, La Jolla, CA 92093, USA}

\author[0000-0003-4714-3829]{Nikolaj Stevnbak Andersen}
\affil{Backyard Worlds: Planet 9}

\author[0000-0001-8662-1622]{Frank Kiwy}
\affil{Backyard Worlds: Planet 9}

\author[0000-0001-5284-9231]{Melina Th{\'e}venot}
\affil{Backyard Worlds: Planet 9}

\author{The Backyard Worlds: Planet 9 Collaboration}

\date{January 2022}

\begin{abstract}
While stars are often found in binary systems, brown dwarf binaries are much rarer. Brown dwarf--brown dwarf pairs are typically difficult to resolve because they often have very small separations. Using  brown dwarfs discovered with data from the Wide-field Infrared Survey Explorer (WISE) via the Backyard Worlds: Planet 9 citizen science project, we inspected other, higher resolution, sky surveys for overlooked cold companions. During this process we discovered the brown dwarf binary system CWISE J0146$-$0508AB, which we find has a very small chance alignment probability based on the similar proper motions of the components of the system.  Using follow-up near-infrared spectroscopy with Keck/NIRES, we determined component spectral types of L4 and L8 (blue), making CWISE J0146$-$0508AB one of only a few benchmark systems with a blue L dwarf. At an estimated distance of $\sim$40 pc, CWISE J0146$-$0508AB has a projected separation of $\sim$129 AU, making it the widest separation brown dwarf pair found to date. We find that such a wide separation for a brown dwarf binary may imply formation in a low-density star-forming region.  
\end{abstract}

\section{Introduction}
\label{sec:introduction}

Several of the first brown dwarf discoveries were part of binary systems with a stellar component, such as GD 165 B \citep{becklin1988} and Gl 229B \citep{Nakajima1995}. The first brown dwarf binary systems, defined here as systems where both components are below the hydrogen burning minimum mass ($\sim$0.07 $M_{\odot}$; \citealt{Saumon2008}), required the use of high-resolution imaging (e.g., \citealt{Martin1998, martin1999}) or high-resolution spectroscopic monitoring (e.g., \citealt{Basri1999}).  As with the binary fraction for stars, the brown dwarf binary fraction can help to put constraints on substellar formation theories (e.g., \citealt{Bate2009}). The stellar multiplicity fraction decreases significantly with decreasing primary mass, from $\sim$69\% for A-type stars \citep{DeRosa2014} to $\sim$25--30\% for M-type stars \citep{ward2015, Winters2019}.  This decreasing binary fraction continues across the stellar--substellar boundary, with measured brown dwarf binary fractions typically between $\sim$10--20\% (e.g., \citealt{Burgasser2006, Radigan2013, Aberasturi2014, Fontanive2018}).  Further, the semi-major axis distribution decreases significantly for substellar objects, with brown dwarf - brown dwarf separations typically of a few AU (e.g., \citealt{Artigau2011, Faherty2020}). 

There are $\sim$50 resolved, field age brown dwarf binaries \citep{Bardalez2015, Faherty2020}, and their median separation is $\approx$4 AU.  Only four systems have separations wider than 20 AU (one of which is a hierarchical triple; \citealt{Radigan2013}), with the widest system known to date being SDSS J1416+1348AB \citep{Burgasser2010B, Burningham2010, Scholz2010}, with a projected separation of $\sim$89 AU.  There are several known brown dwarf pairs with wider separations at very young ages (e.g., \citealt{Chauvin2004, Luhman2004, Close2007, Bejar2008, Luhman2009, Strampelli2020, DeFurio2021}), though some of these systems may still be disrupted, depending on the density of their birth environment.  Note that some very wide L+L and L+T binaries exist (e.g., \citealt{Faherty2020, Marocco2020}), but in these cases the L-type component is most likely stellar. 

Brown dwarf binaries are most commonly found through either spectral decomposition (e.g., \citealt{Burgasser2010, Geissler2011, Bardalez2014, Marocco2015}) or high resolution imaging (e.g., \citealt{Reid2001, Bouy2003, Burgasser2003, Gizis2003, Burgasser2006, Liu2006, Reid2006, Dupuy2012, Aberasturi2014, Bardalez2015, Opitz2016}), with a small number of low-mass systems found through radial velocity (e.g., \citealt{Basri1999, Blake2008}) or astrometric (e.g. \citealt{Sahlmann2013}) monitoring. Until now, only one field brown dwarf binary has been discovered as a resolved pair in wide-field survey images -- the previously mentioned SDSS J1416$+$1348AB, which was resolved in UKIDSS images \citep{Burgasser2010B, Burningham2010, Scholz2010}.  

In this paper we present the discovery of the wide brown dwarf binary CWISE J014611.20$-$050850.0AB. We describe its discovery in Section \ref{sec:discovery} and our follow-up spectroscopic observations in Section \ref{sec:obs}.  We present our analysis and discussion of the system in Sections \ref{sec:analysis} and \ref{sec:discussion}.    

\section{Discovery}
\label{sec:discovery}
The Backyard Worlds: Planet 9 (BYW) citizen science project is a collaborative effort between professional astronomers and citizen scientists from around the world \citep{Kuchner2017}. One of the main goals of the project is to create a more complete census of substellar members of the solar neighborhood.  The BYW project has so far made numerous contributions to the census of solar neighborhood members (e.g., \citealt{Bardalez2020, Meisner2020, Kirkpatrick2021, Schneider2021}). The project has also been adept at discovering cold companions to nearby stars (e.g., \citealt{Faherty2020, Jalowiczor2021, Rothermich2021, Faherty2021}). 

\begin{figure*}
\centering\includegraphics[scale=.5]{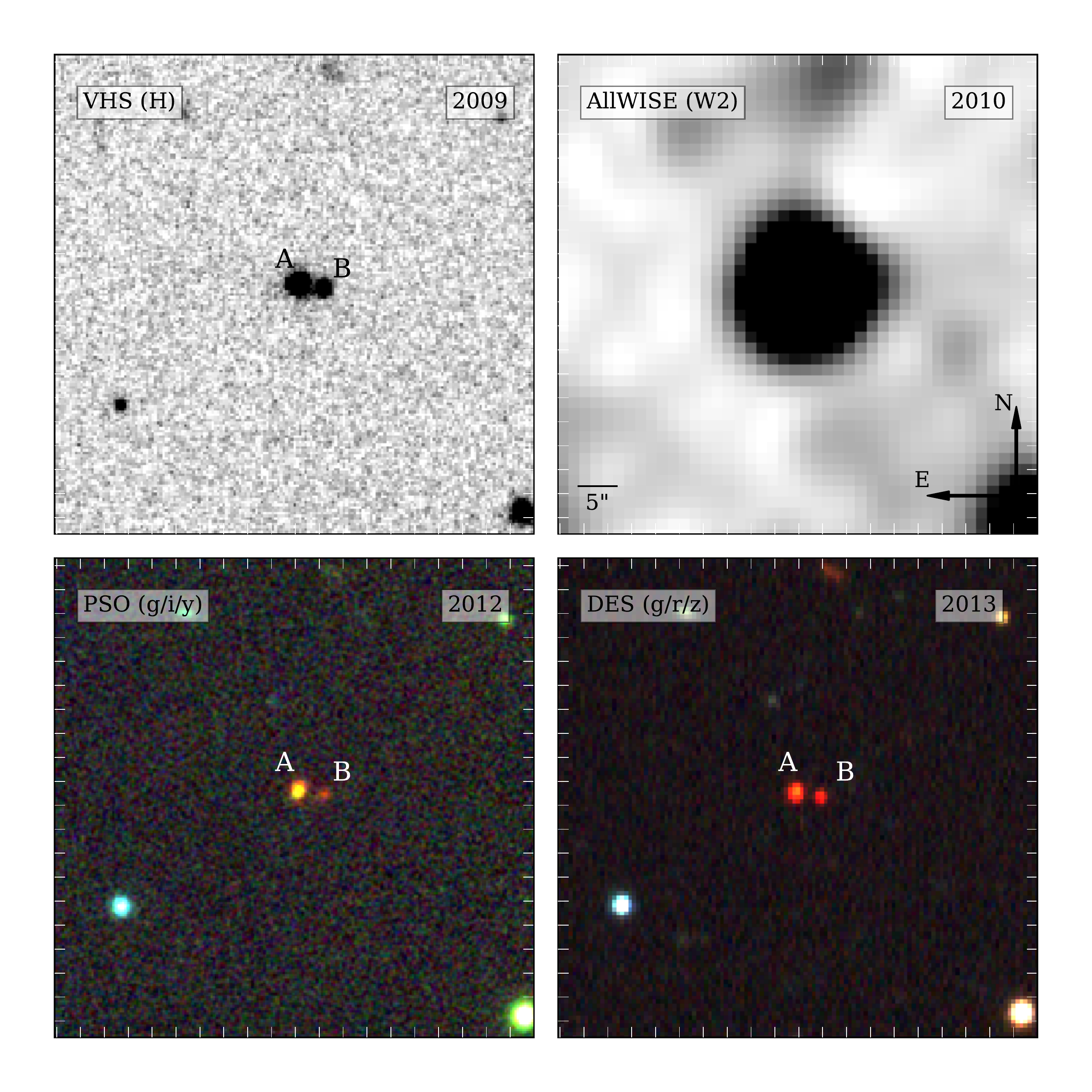}
\caption{Survey images of CWISE J0146$-$0508AB. From top left to bottom right; VHS ($H$), AllWISE (W2), Pan-STARRS1 3-color ($g$/$i$/$y$), and DES 3-color ($g$/$r$/$z$). While it it difficult to resolve CWISE J0146$-$0508AB in the AllWISE image, it is clearly resolved as a double in the VHS, Pan-STARRS, and DES images.  Note also that the reddish colors of each component in the Pan-STARRS1 and DES images indicates very red optical colors, typical of cold brown dwarfs.  Each image is 1\arcmin\ $\times$ 1\arcmin. }
\label{fig:finder}
\end{figure*}

The BYW project uses data from the Wide-field Infrared Survey Explorer (WISE; \citealt{Wright2010}) to identify brown dwarf candidates.  In an effort to identify previously unknown brown dwarf binaries, we have examined each brown dwarf candidate found through the BYW project for evidence of binarity in existing imaging surveys with higher resolution than WISE.  These surveys include Pan-STARRS1 \citep{Chambers2016}, the VISTA Hemisphere Survey (VHS; \citealt{McMahon2013}), the UKIRT Hemisphere Survey (UHS; \citealt{Dye2018}), and the Dark Energy Survey (DES; \citealt{DES2005}). This search of over 3000 brown dwarf candidates returned $\sim$10 candidate binaries based on proximity, available colors, and proper motions.  We have obtained follow-up spectroscopy for both components of one of the most promising binary candidates, CWISE J014611.20$-$050850.0A and CWISE J014611.20$-$050850.0B (\citealt{Marocco2020}; CWISE J0146$-$0508AB hereafter), which was prioritized for follow-up observations because it had clear detections in multiple surveys (Figure \ref{fig:finder}). The remaining brown dwarf candidates will need similar follow-up observations to confirm their binary status.  

CWISE J0146$-$0508A was submitted as an object of interest by citizen scientists Nikolaj Stevnbak, Sam Goodman, Melina Th{\'e}venot, Dan Caselden, and Frank Kiwy based on its significant proper motion.  The inspection of archival images of CWISE J0146$-$0508A from Pan-STARRS1, VHS, and DES revealed a possible resolved companion at a separation of $\sim$3\arcsec (Figure \ref{fig:finder}). While the CatWISE 2020 catalog has an entry at the position of the putative secondary, the measured proper motions from the CatWISE 2020 catalog are significantly different than CWISE J0146$-$0508A, likely due to the sources being blended in the {\it WISE} images.  The positions and photometry from each survey for both components of the pair are listed in Table 1.

\begin{deluxetable*}{lcccc}
\label{tab:W0146AB}
\tablecaption{Properties of CWISE J0146$-$0508AB Components}
\tablehead{
\colhead{Parameter} & \colhead{CWISE J0146$-$0508A} & \colhead{CWISE J0146$-$0508B}}
\startdata
\cutinhead{This Work}
$\mu_{\alpha}$ (mas yr$^{-1}$) & 79.14$\pm$1.10 & 83.94$\pm$3.43 \\	
$\mu_{\delta}$ (mas yr$^{-1}$) & $-$214.40$\pm$1.03 & $-$210.33$\pm$3.35 \\
Spec.\ Type & L4 & L8 (blue) \\
$T_{\rm eff}$ (K) & 1720$\pm$150 & 1340$\pm$140 \\
Mass ($M_{\rm Jup}$)\tablenotemark{a} & 72$\pm$6 & 66$\pm$10 \\
\cutinhead{CatWISE 2020}
R.A. (J2000) ($\degr$) & 26.5467046 & 26.5458109 \\
Dec. (J2000) ($\degr$) & $-$5.1472375 & $-$5.1473468 \\
W1 (mag) & 13.779$\pm$0.028 & 15.069$\pm$0.027 	\\
W2 (mag) & 13.507$\pm$0.029 & 14.753$\pm$0.029 \\
\cutinhead{Pan-STARRS1 DR2}
R.A. (J2000) ($\degr$) & 26.54666153 & 26.54576435 \\
Dec. (J2000) ($\degr$) & $-$5.14708987 & $-$5.1472229 \\
$i$ (mag) & 20.692$\pm$0.020 & $\dots$ \\
$z$ (mag) & 19.125$\pm$0.019 & 20.476$\pm$0.032 \\
$y$ (mag) & 18.074$\pm$0.012 & 19.379$\pm$0.124 \\
\cutinhead{DES DR1}
R.A. (J2000) ($\degr$) & 26.546694 & 26.545828 \\
Dec. (J2000) ($\degr$) &	$-$5.147201 & $-$5.147353 \\
$i$ (mag) & 20.136$\pm$0.011 & 22.089$\pm$0.061 \\
$z$ (mag) & 18.588$\pm$0.006 & 20.055$\pm$0.020 \\
$Y$ (mag) & 18.057$\pm$0.008 & 19.572$\pm$0.028 \\
\cutinhead{VHS DR6}
R.A. (J2000) ($\degr$) & 26.5465517 & 26.5456781 \\
Dec. (J2000) ($\degr$) & $-$5.1469086 & $-$5.147062 \\
$J$ (mag) & 15.818$\pm$0.005 & 17.420$\pm$0.018 \\
$H$ (mag) & 15.044$\pm$0.006 & 16.524$\pm$0.019\\
$K_S$ (mag) & 14.347$\pm$0.006 & 15.790$\pm$0.021\\
\enddata
\tablenotetext{a}{Estimated using the \cite{Phillips2020} evolutionary models.}
\end{deluxetable*}

\section{Observations}
\label{sec:obs}

\subsection{Keck/NIRES}
To measure the spectral types of CWISE J0146$-$0508A and CWISE J0146$-$0508B, we observed both components with the Near-Infrared Echellette Spectrometer (NIRES; \citealt{Wilson2004}) located on the Keck II telescope on UT 22 October 2020. NIRES provides R$\sim$2,700 spectra from 0.9--2.45 $\mu$m.  For each component, eight 300 s exposures were taken for total on-source integration times of 2400 s.  Each target was nodded along the slit in an ABBA pattern. The A0 star HD 216807 was observed for telluric correction purposes.  The spectra were extracted using a modified version of the SpeXTool package \citep{Vacca2003, Cushing2004}. We achieved a S/N of 400 and 1000 for the J and K peaks of the A component, respectively, and a S/N of 50 and 200 for the J and K peaks of the B component. The final reduced spectra are shown in Figure 2.

\begin{figure*}
\plotone{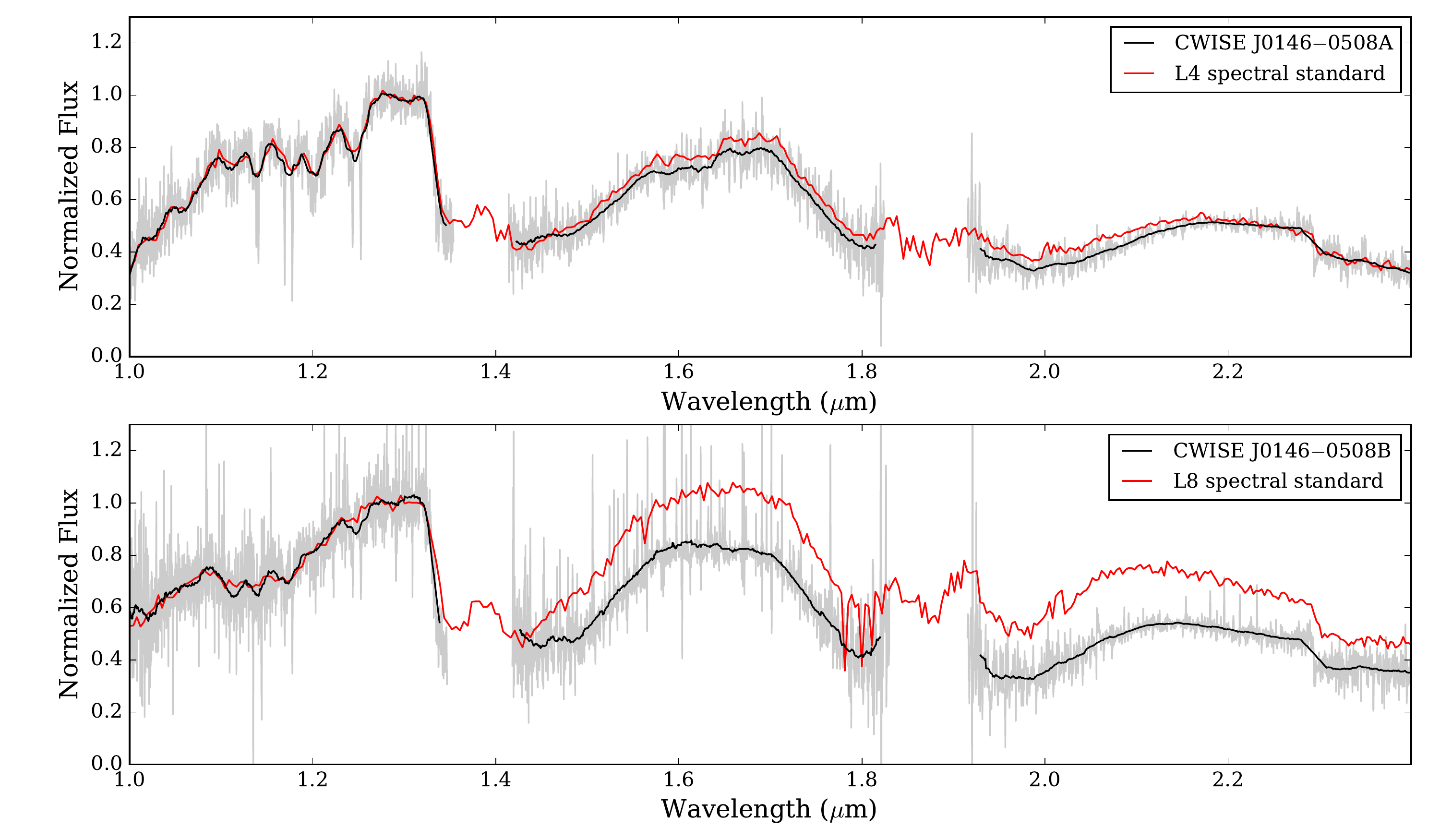}
\caption{Keck/NIRES spectra of CWISE J0146$-$0508A (top) and CWISE J0146$-$0508B (bottom) compared to the L4 (2MASS J21580457$-$1550098; \citealt{kirkpatrick2010}) and L8 (2MASSW J1632291+190441; \citealt{Burgasser2007}) spectral standards following \cite{kirkpatrick2005}. The spectra of CWISE J0146$-$0508A and CWISE J0146$-$0508B are shown at full resolution (grey lines) and smoothed to the resolution of the spectral standards (black lines). All spectra have been normalized between 1.27 and 1.29 $\mu$m.  }
\label{fig:spectra}
\end{figure*}

\section{Analysis}
\label{sec:analysis}

\subsection{Spectral Types}
Spectral types for each component were determined by comparing $J-$band morphologies to near-infrared spectral standards from \cite{Burgasser2006B} and \cite{kirkpatrick2010}. We found a best-fit type of L4 $\pm$0.5 subtypes for the A component and L8 $\pm$ 0.5 subtypes for the B component (Figure \ref{fig:spectra}). CWISE J0146$-$0508B appears bluer than the L8 spectral standard, and we therefore give it a spectral type designation of L8 (blue). Neither spectrum shows signs of being a spectral binary according to the spectral indices of \cite{Bardalez2014,Bardalez2015}.  

\subsection{Proper Motion and Chance Alignment Probability}

CWISE J0146$-$0508A has an entry in the {\it Gaia} EDR3 catalog \citep{Gaia2021}, but includes no measured proper motion or parallax values.  This is likely because CWISE J0146$-$0508A has a {\it Gaia} $G-$magnitude of 20.99, which is fainter than the nominal $G$ = 20.7 mag survey limit.

Even though CWISE J0146$-$0508A and CWISE J0146$-$0508B are resolved in the CatWISE2020 catalog, they are blended in {\it WISE} images (Figure \ref{fig:finder}), making their CatWISE measured proper motions potentially unreliable.  For this reason, we calculated the proper motion of each component using available astrometric measurements for which this pair is well-resolved.  We used single-epoch positions of each component from the NOIRLab Source Catalog (NSC) DR2 \citep{nidever2021} and Pan-STARRS1 DR2 \citep{Flewelling2018}.  The NSC contains single-epoch measurements from the Dark Energy Camera (DECam) on the CTIO-4 m.  For CWISE J0146$-$0508A, there are 31 single-epoch NSC DR2 detections and 55 single-epoch Pan-STARRS1 DR2 detections.  For CWISE J0146$-$0508B, there are 22 single-epoch NSC DR2 detections and 9 single-epoch Pan-STARRS1 DR2 detections.  For both objects, these positional measurements span a range of just over 9 years, from 2009 to 2018.  There are significantly fewer Pan-STARRS1 DR2 detections of CWISE J0146$-$0508B than CWISE J0146$-$0508A because CWISE J0146$-$0508B was not detected with the $i$-band filter in any epoch (CWISE J0146$-$0508A has 19 $i$-band detections), and is near the 5$\sigma$ Pan-STARRS1 single-epoch detection limits of 20.9 mag at $z$-band and 19.7 mag at $y$-band \citep{Chambers2016}.

We calculated proper motions using a least-squares fit weighted by the quoted uncertainty of each astrometric measurement.  For CWISE J0146$-$0508A, we find $\mu_{\alpha}$ = 79.14$\pm$1.10 mas yr$^{-1}$ and $\mu_{\delta}$ = $-$214.40$\pm$1.03 mas yr$^{-1}$.  For CWISE J0146$-$0508B, we find $\mu_{\alpha}$ = 83.94$\pm$3.43 mas yr$^{-1}$ and $\mu_{\delta}$ = $-$210.33$\pm$3.35 mas yr$^{-1}$.  These values are fully consistent with the proper motion values given in the NSC DR2 for CWISE J0146$-$0508A ($\mu_{\alpha}$ = 81.79$\pm$1.98 mas yr$^{-1}$, $\mu_{\delta}$ = $-$218.51$\pm$1.93 mas yr$^{-1}$) and CWISE J0146$-$0508B ($\mu_{\alpha}$ = 80.52$\pm$4.85 mas yr$^{-1}$, $\mu_{\delta}$ = $-$214.01$\pm$4.75 mas yr$^{-1}$).

The small separation of this pair ($\sim$3\arcsec) and similarity of proper motion components (within $\sim$1$\sigma$) suggests that this is likely a physical pair.  We use the \texttt{CoMover} code \citep{Gagne2021, Schneider2021} to evaluate the possibility that this pair is the result of a chance alignment.  We find a $>$99$\%$ probability that this is a physically associated pair. 

\subsection{Distance/Separation}
\label{sec:dist}

\cite{Kirkpatrick2021} provides relations between absolute magnitude and spectral type for L, T, and Y type brown dwarfs within 20 pc.  Using their relations for $J$ and $H$, we find a distance of 41.5$\pm$2.5 pc for CWISE J0146$-$0508A and 40.0$\pm$3.0 pc for CWISE J0146$-$0508B using weighted averages for $J$ and $H$.  We give a conservative estimate for the distance of the pair of 41$\pm$5 pc.    

The average separation between CWISE J0146$-$0508A and CWISE J0146$-$0508B from NSC DR2 and Pan-STARRS1 DR2 measurements is 3\farcs14$\pm$0\farcs8. Using our distance estimate and measured separation we find a projected separation of 129$\pm$15 AU for this pair.  We account for inclination angle and eccentricity effects when converting to a physical separation following \cite{Dupuy2011} and find a separation of 149$^{+104}_{-41}$ AU.

\subsection{Physical Properties (Effective Temperature, Age, and Mass)}
\label{sec:props}

Using our derived spectral types and a spectral type uncertainty of $\pm$0.5 subtypes, we find effective temperature ($T_{\rm eff}$) values of 1720$\pm$150 K and 1340$\pm$140 K for CWISE J0146$-$0508A and CWISE J0146$-$0508B, respectively, using the spectral type vs.\ $T_{\rm eff}$ relation from \cite{Kirkpatrick2021}.

Neither CWISE J0146$-$0508A nor CWISE J0146$-$0508B show any signs of youth in their spectra, as a low surface gravity will typically show weakened absorption features from FeH and alkali lines, and stronger absorption from VO \citep{allers2013}. We find a gravity classification of FLD-G for CWISE J0146$-$0508A using the classification scheme of \cite{allers2013}. The \cite{allers2013} gravity classification method does not extend to L8 spectral types, so is not applicable to CWISE J0146$-$0508B.  We note that most young, late-L type brown dwarfs are significantly redder than field age standards (e.g., \citealt{gizis2012, faherty2013, liu2013, schneider2014, kellogg2016, schneider2016}). CWISE J0146$-$0508B, on the other hand, is bluer than the L8 standard (Figure \ref{fig:spectra}).  This is also seen in the measured $J-K$ color of CWISE J0146$-$0508B (1.63 mag), which is on the blue end of the field-age L8 $J-K$ color distribution ($\sim$1.78$\pm$0.16 mag; \citealt{Faherty2016}).  A blue color is generally attributed to thin condensate clouds and/or old age (e.g., \citealt{Cushing2010, Burgasser2010B}) or perhaps viewing geometry \citep{vos2017}.  CWISE J0146$-$0508B is therefore a compelling benchmark target for clarifying the nature of the blue L dwarf population. These characteristics imply that the CWISE J0146$-$0508AB system is not young.

While a blue near-infrared color is a trait common to low-metallicity, old ($\gtrsim$10 Gyr) subdwarfs, CWISE J0146$-$0508B is not nearly as blue as the sdL7 standard from \cite{Greco2019}. Brown dwarfs with spectral types of sdL7 or sdL8 in \cite{zhang2018} have $J-K$ colors between 0.9 and 1.3 mag, much bluer than that found for CWISE J0146$-$0508B (1.63 mag). Further, CWISE J0146$-$0508A is a good match to the L4 standard at all wavelengths.  Using the proper motion we derived for CWISE J0146$-$0508A and our estimated distance from Section \ref{sec:dist}, we find a tangential velocity for this system of $\sim$44 km s$^{-1}$, consistent with the thin disk population (e.g. \citealt{Nissen2004}).  We therefore conclude that CWISE J0146$-$0508A and CWISE J0146$-$0508B are not subdwarfs.  We assume a conservative age range of 0.5--10 Gyr for this pair.

Using our derived $T_{\rm eff}$ values and this age range, we find a mass of 72$\pm$6 $M_{\rm Jup}$ for CWISE J0146$-$0508A and a mass of 66$\pm$10 $M_{\rm Jup}$ for CWISE J0146$-$0508B using the evolutionary models of \cite{Phillips2020}.  Uncertainties are determined in a Monte Carlo fashion, using a uniform probability distribution for age over the range 0.5--10 Gyr and a normal distribution for $T_{\rm eff}$. We also calculated masses using the hybrid evolutionary models of \cite{Saumon2008} and find fully consistent results of 74$\pm$5 $M_{\rm Jup}$ and 64$\pm$10 $M_{\rm Jup}$ for CWISE J0146$-$0508A and CWISE J0146$-$0508B, respectively.  While the current mass estimate for CWISE J0146$-$0508A suggests there is a possibility it is stellar and not a brown dwarf, previous studies of the location of the stellar/substellar boundary have found it to occur at a spectral type of $\approx$L4 or earlier \citep{Dieterich2014, Dupuy2017}, making CWISE J0146$-$0508A most likely substellar in nature.  More precise mass estimates for each component will  require a tighter age constraint.  

\section{Discussion}
\label{sec:discussion}

\subsection{Binding Energy}
Using the estimated separation (149$^{+104}_{-41}$ AU) and the masses of CWISE J0146$-$0508A (72$\pm$6 $M_{\rm Jup}$) and CWISE J0146$-$0508B (66$\pm$10 $M_{\rm Jup}$), we calculate the binding energy of this system.  We find a binding energy of 4.0$^{+0.9}_{-0.7}$ $\times$10$^{41}$ ergs.  This is one of the lowest binding energies found for field age brown dwarf--brown dwarf pairs \citep{Faherty2020}.  

Previous investigations have found that wide brown dwarf binaries with separations greater than a few tens of AU are exceptionally rare, especially at field ages. There have been several efforts to describe the largest possible separations for low-mass systems, both empirically \citep{Reid2001, Burgasser2003, Close2003, Dhital2010} and theoretically \citep{Zuckerman2009, Faherty2010}. Specifically for brown dwarf binaries, \cite{Burgasser2003} showed that being disrupted either through chance encounters with a star or giant molecular cloud were unlikely, and thus dissipation due to random encounters could not explain the lack of wide brown dwarf binaries.  \cite{Close2007} performed a similar investigation considering birth environments, which they estimated to be several thousand times more dense than the typical Galactic field.  Therefore a star or brown dwarf would have a much higher probability of chance encounters in its higher density birth environment. They found that 10 Myr in a dense cluster environment would be much more disruptive for a binary brown dwarf than 10 Gyr in the field. They suggested that the known, young ($\lesssim$10 Myr), wide ($\gtrsim$20 AU) brown dwarf binaries are therefore in the process of being disrupted.  This would explain why wide brown dwarf binaries have been found in young clusters, but very few have been found at field ages.  However, we note here that the cluster density used by \cite{Close2007} ($n_*$ $\sim$1000 pc$^{-3}$) is likely an underestimate for the densest star-forming regions like the Orion Nebula Cluster (ONC) ($n_*$ $\sim$3700 pc$^{-3}$; \citealt{DeFurio2021}) and a significant overestimate for low-density regions such as Taurus ($n_*$ $\sim$1--10 pc$^{-3}$; \citealt{Luhman2018}).  This is further complicated by the fact that cluster densities change as the cluster ages (e.g., \citealt{Parker2014}). 

\begin{figure*}
\plotone{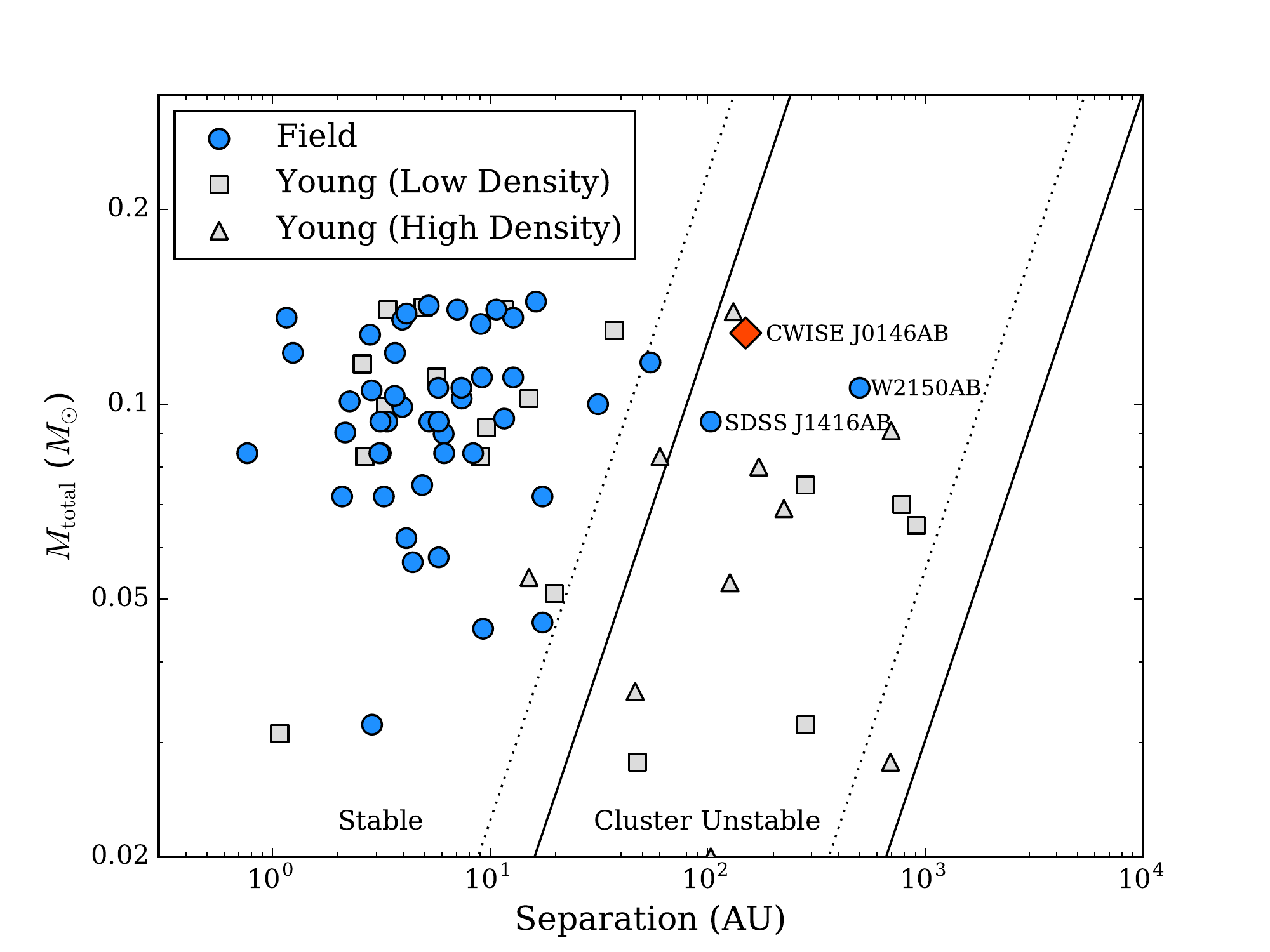}
\caption{Separation versus total mass for all resolved binaries from  \cite{Faherty2020}, \cite{Strampelli2020} and \cite{DeFurio2021} where each component has a mass less than $<$0.075 $M_{\odot}$.  Field age pairs are represented by blue circles, while young binaries from dense regions (e.g., ONC) are shown as gray triangles and young binaries from low-density regions (e.g., Taurus, Chamaeleon, $\rho$ Ophiucus) are shown as gray squares.  Dotted and solid lines show the instability regions defined in \cite{Close2007}, with dotted lines representing the maximum bound separations for cluster (left) and field (right) due to diffusion and solid lines representing maximum separations due to close, catastrophic encounters. These regions are most relevant to the young objects in high-density star-forming regions.}
\label{fig:separation}
\end{figure*}

Figure \ref{fig:separation} shows the projected separations and total masses for all resolved binaries from \cite{Faherty2020}, with each component having a mass near the substellar boundary ($<$0.075 $M_{\odot}$; \citealt{Saumon2008}).  All projected separations from \cite{Faherty2020} are multiplied by 1.16 following \cite{Dupuy2011}. To this list we add the recent substellar binaries discovered in the ONC from \cite{Strampelli2020} and \cite{DeFurio2021}. \cite{Close2007} defined instability regions based on gradual tidal diffusion and close catastrophic encounters for field and cluster densities.  \cite{Close2007} suggested that the young brown dwarf binaries in the ``Cluster Unstable'' region of Figure \ref{fig:separation} will likely dissipate before joining the field.  Since the study of \cite{Close2007}, there are now three field age binaries that reside in the ``Cluster Unstable'' region, CWISE J0146$-$0508AB, SDSS J1416+1348AB \citep{Burgasser2010B, Burningham2010, Scholz2010}, and WISE 2150$-$7520AB \citep{Faherty2020}. While the A component of WISE 2150$-$7520AB, with a spectral type of L1, is likely stellar \citep{Dieterich2014, Dupuy2017}, it is just above the stellar/substellar boundary, and thus its survival as a system is similarly puzzling as with CWISE J0146$-$0508AB and SDSS J1416$+$1348AB.   

As discussed in \cite{Luhman2012} and \cite{Todorov2014}, while most stars and brown dwarfs likely form in high density environments like the ONC where wide brown dwarf binary survival is unlikely, some fraction form in lower-density regions like Taurus \citep{Luhman2009} and Chamaeleon \citep{Luhman2004} where the probability of survival is much higher.  It may be that systems such as CWISE J0146$-$0508AB, SDSS J1416$+$1348AB, and WISE 2150$-$7520AB formed in such a relatively low-density region, allowing them to have survived as low-mass binaries at such wide separations to field ages.  

Resolved brown dwarf binaries are often found to be valuable as benchmarks for measuring dynamical masses from orbital monitoring (e.g., \citealt{Konopacky2010, Dupuy2017}). Assuming our measured separation is the true semi-major axis of the system, and using the masses found in Section \ref{sec:props}, we find an orbital period of $\sim$5000 years.  Therefore the $\sim$9 years of observations so far have covered less than 0.2\% of an orbital period.  While it has recently been shown that accurate orbital parameters can be produced with only a small fraction of the orbit having been monitored (e.g., \citealt{Blunt2017}), obtaining a fraction of CWISE J0146$-$0508AB's orbit greater than a few percent would be beyond our lifetimes.  

\subsection{Future Work}
The vast majority of the current census of known field brown dwarfs were discovered with wide-field infrared surveys such as 2MASS or WISE. Subsequently, new, higher resolution surveys have been produced, such as the Vista Hemisphere Survey (VHS), the UKIRT Hemisphere Survey (UHS), Pan-STARRS1, and the Dark Energy Survey (DES). The process demonstrated in this paper shows the value of reexamining known or suspected brown dwarf candidates for evidence of close companions.  It is possible that the examination of images from these surveys for previously missed companions to known brown dwarfs may prove similarly fruitful. By expanding this analysis to all confirmed brown dwarfs we would be able to  develop a more comprehensive census of exceptionally wide brown dwarf binary systems. Such a survey would be complementary to focused high resolution imaging surveys (e.g., \citealt{Bardalez2015, Fontanive2018}) and other binary identification techniques (e.g., \citealt{Deacon2017}) that may not be sensitive to such wide companions.  Further, the Roman Space Telescope \citep{Spergel2015} and Euclid \citep{Laureijs2011} future space missions will have {\it HST}-like resolution in the near-infrared over large areas of the sky, allowing for brown dwarf binary surveys at much smaller separations than is currently achievable with ground-based imaging surveys.

Such binaries have the potential to be used as valuable benchmark systems for atmospheric retrievals, which can determine chemical abundances, metallicities, and temperature-pressure profiles (e.g., \citealt{Line2015, Burningham2017, Line2017, Zalesky2019, Gonzales2020}).  Presumably, CWISE J0146$-$0508A and CWISE J0146$-$0508B formed out of the same cloud of gas and dust, and should therefore have similar elemental abundances.  This pair can then be used as a test for such retrieval methods.  And, because CWISE J0146$-$0508A has a two-parameter (R.A. and Dec.) solution in {\it Gaia} EDR3, we can expect a full five-parameter solution (R.A., Dec., parallax, and proper motion components) in a future {\it Gaia} release \citep{Fabricius2021}.  As a benchmark system containing a blue L dwarf, CWISE J0146$-$0508AB may be able to elucidate the unique atmospheric chemistry of such objects and help to probe cloud properties across the L/T transition \citep{Brock2021}.

\facilities{Keck (NIRES)}

\software{SpeXTool \citep{Cushing2010}; SPLAT \citep{Burgasser2017}; CoMover \citep{Gagne2021} }

\section{Acknowledgements}
We thank the anonymous referee for comments that improved this paper. The Backyard Worlds: Planet 9 team would like to thank the many Zooniverse volunteers who have participated in this project. We would also like to thank the Zooniverse web development team for their work creating and maintaining the Zooniverse platform and the Project Builder tools. This research was supported by NASA grant 2017-ADAP17-0067. This material is based upon work supported by the National Science Foundation under Grant No. 2007068, 2009136, and 2009177.  

Some of the data presented herein were obtained at the W. M. Keck Observatory, which is operated as a scientific partnership among the California Institute of Technology, the University of California and the National Aeronautics and Space Administration. The Observatory was made possible by the generous financial support of the W. M. Keck Foundation.  This publication makes use of data products from the {\it Wide-field Infrared Survey Explorer}, which is a joint project of the University of California, Los Angeles, and the Jet Propulsion Laboratory/California Institute of Technology, and NEOWISE which is a project of the Jet Propulsion Laboratory/California Institute of Technology. {\it WISE} and NEOWISE are funded by the National Aeronautics and Space Administration.  The authors wish to recognize and acknowledge the very significant cultural role and reverence that the summit of Maunakea has always had within the indigenous Hawaiian community.  We are most fortunate to have the opportunity to conduct observations from this mountain.  

The Pan-STARRS1 Surveys (PS1) and the PS1 public science archive have been made possible through contributions by the Institute for Astronomy, the University of Hawaii, the Pan-STARRS1 Project Office, the Max-Planck Society and its participating institutes, the Max Planck Institute for Astronomy, Heidelberg and the Max Planck Institute for Extraterrestrial Physics, Garching, The Johns Hopkins University, Durham University, the University of Edinburgh, the Queen's University Belfast, the Harvard-Smithsonian Center for Astrophysics, the Las Cumbres Observatory Global Telescope Network Incorporated, the National Central University of Taiwan, the Space Telescope Science Institute, the National Aeronautics and Space Administration under Grant No. NNX08AR22G issued through the Planetary Science Division of the NASA Science Mission Directorate, the National Science Foundation Grant No. AST-1238877, the University of Maryland, Eotvos Lorand University (ELTE), the Los Alamos National Laboratory, and the Gordon and Betty Moore Foundation.

\bibliography{bibliography.bib}

\end{document}